\documentclass[fleqn,usenatbib]{mnras}
\usepackage[T1]{fontenc}
\usepackage{ae,aecompl}
\usepackage{graphicx}
\usepackage{epstopdf}
\usepackage{amsmath}
\usepackage{amssymb}
\usepackage{enumerate}
\usepackage{mathabx}

\pubyear{2020}
\begin{document}

\def\func#1{\mathop{\rm #1}\nolimits}
\def\unit#1{\mathord{\thinspace\rm #1}}

\title[Braking index and PWN emission of PSR B0540-69]{The braking index of
PSR B0540-69 and the associated pulsar wind nebula emission after spin-down
rate transition}
\author[Wang et al.]{ L. J. Wang$^{1}$\thanks{%
E-mail: wanglingjun@ihep.ac.cn}, M. Y. Ge$^{1}$, J. S. Wang$^{2}$\thanks{%
E-mail: jiesh.wang@gmail.com}, S. S. Weng$^{3}$, H. Tong$^{4}$, L. L. Yan$%
^{5}$, \newauthor S. N. Zhang$^{1,6}$, Z. G. Dai$^{7,8}$, L. M. Song$^{1,6}$ 
\\
$^{1}$ Key Laboratory of Particle Astrophysics, Institute of High Energy
Physics, Chinese Academy of Sciences, Beijing 100049, China\\
$^{2}$ Tsung-Dao Lee Institute, Shanghai Jiao Tong University, Shanghai
200240, China\\
$^{3}$ Department of Physics and Institute of Theoretical Physics, Nanjing
Normal University, Nanjing 210023, China \\
$^{4}$ School of Physics and Electronic Engineering, Guangzhou University,
510006 Guangzhou, China\\
$^{5}$ School of Mathematics and Physics, Anhui Jianzhu University, Hefei
230601, China \\
$^{6}$ University of Chinese Academy of Sciences, Chinese Academy of
Sciences, Beijing 100049, China\\
$^{7}$ School of Astronomy and Space Science, Nanjing University, Nanjing
210093, China \\
$^{8}$ Key Laboratory of Modern Astronomy and Astrophysics (Nanjing
University), Ministry of Education, Nanjing 210093, China}
\date{Accepted XXX. Received YYY; in original form ZZZ}
\maketitle

\begin{abstract}
In Dec. 2011 PSR B0540-69 experienced a spin-down rate transition (SRT),
after which the spin-down power of the pulsar increased by $\sim 36\%$.
About 1000 days after the SRT, the X-ray luminosity of the associated pulsar
wind nebula (PWN) was found to brighten by $32\pm 8\%$. After the SRT, the
braking index $n$ of PSR B0540-69 changes from $n=2.12$ to $n=0.03$ and then
keeps this value for about five years before rising to $n=0.9$ in the
following years. We find that most of the current models have difficulties
in explaining the measured braking index. One exceptive model of the braking
index evolution is the increasing dipole magnetic field of PSR B0540-69. We
suggest that the field increase may result from some instabilities within
the pulsar core that enhance the poloidal component at the price of toroidal
component of the magnetic field. The increasing dipole magnetic field will
result in the X-ray brightening of the PWN. We fit the PWN X-ray light curve
by two models: one assumes a constant magnetic field within the PWN during
the brightening and the other assumes an enhanced magnetic field
proportional to the energy density of the PWN. It appears that the two
models fit the data well, though the later model seems to fit the data a bit
better. This provides marginal observational evidence that magnetic field in
the PWN is generated by the termination shock. Future high-quality and
high-cadence data are required to draw a solid conclusion.
\end{abstract}

\label{firstpage} \pagerange{\pageref{firstpage}--\pageref{lastpage}}


\section{Introduction}

An isolated pulsar loses its rotational energy to relativistic particles and
electromagnetic radiation. This process reduces the spin frequency $\nu $ of
the pulsar. The spin-down rate $\dot{\nu}$ of the pulsar is usually modelled
as $\dot{\nu}=-\kappa \nu ^{n}$, where $\kappa $ is related to the
energy-loss mechanisms of the neutron star and $n=\ddot{\nu}\nu /\dot{\nu}%
^{2}$ is the braking index. Till now, the braking indices of nine young
pulsars have been measured \citep{Lyne15, Archibald16,
Ou16} and these measurements show that usually $0<n<3$, with one exception
for PSR J1640--4631, which has a braking index $n=3.15$ \citep{Archibald16}.

PSR B0540-69 is a 1100-year-old young pulsar located in the Large Magellanic
Cloud with a spin period of $50\unit{ms}$ \citep{Mathewson1980,Seward1984}.
The spin-down of PSR B0540-69 is relatively stable before Dec. 2011. Using
15.8 years of data from the Rossi X-ray Timing Explorer, \cite{Ferdman15}
determined a braking index of $n=2.129\pm 0.012$. In Dec. 2011, PSR B0540-69
experienced a spin-down rate transition (SRT), with $\dot{\nu}$ changing
from $-1.86\times 10^{-10}$ to $-2.52\times 10^{-10}\unit{Hz}\unit{s}^{-1}$
in two weeks, while the spin frequency $\nu $\ is decreasing continuously
and monotonously during and after the SRT \citep{Marshall2015}. In the same
time the braking index also dramatically changed from around 2.12 to 0.031 %
\citep{Marshall2016}, and then increased gradually and slightly to 0.07 in
the following five years after the SRT \citep{Ge2019}. Starting from Dec.
2016 the braking index begins to increase more quickly and in Feb. 2018 $n$
increases to 0.94 \citep{Ge2019}. The SRT is different from a glitch during
which the spin frequency of the pulsar changes suddenly and subsequently
recovers over several weeks to a frequency close to, but not identical with,
that expected by extrapolation from the earlier observations (e.g., %
\citealt{Espinoza11b}; see \citealt{Ge2019} for more detail).

The relativistically out-moving particles (electrons and positrons) emanated
from the central pulsar are decelerated by the termination shock, behind
which particles are accelerated to even higher energy and theoretically
magnetic field should be generated by the shock 
\citep{Pacini1973, Rees1974,
Arons79, Michel94}. As a result, a bright pulsar wind nebula (PWN) is formed
around the pulsar by emitting synchrotron photons 
\citep{Kennel84a, Kennel84b, Coroniti90, Dai04,
Bucciantini11, Wang13, Wang16a}. Accompanied with the SRT, the flux of the
PWN around PSR B0540-69 enhanced gradually after SRT \citep{Ge2019}. It is
shown that this enhancement is correlated to the SRT of the central pulsar %
\citep{Ge2019}.

This paper aims to study the braking index evolution of PSR B0540-69 and the
X-ray brightening of the associated PWN in more detail than in \cite{Ge2019}
and try to explain the evolution theoretically. Besides the data already
used in \cite{Ge2019}, here we also use the data from Insight-HXMT and the
Neutron Star Interior Composition Explorer (NICER). We present the data
reduction in Section \ref{sec:data}. Then we give a plausible interpretation
of the braking index evolution of PSR B0540-69 after SRT in Section \ref%
{sec:n} and then study the PWN brightening in Section \ref{sec:PWN}.
Conclusions and discussion are presented in Section \ref{sec:dis}.

\section{Data reduction}

\label{sec:data}

We select the X-ray telescopes \textit{Insight-HXMT, }\textsl{NuSTAR}, 
\textsl{Swift/XRT} and \textsl{XMM-Newton} to measure the flux of PSR
B0540--69 and its wind nebula. The reduction of \textsl{Swift/XRT} and 
\textsl{NuSTAR} data was presented in \cite{Ge2019}. In this paper we also
include the data collected by \textit{Insight-HXMT}. Launched on June 15th,
2017, \textit{Insight-HXMT }is China's first X-ray astronomical satellite. 
\textit{Insight-HXMT} carries three main instruments: the High Energy X-ray
telescope (HE, $20-250\unit{keV}$, $5000\unit{cm}^{2}$, $1.1^{\degree}\times
5.7^{\degree}$, $\sim 2\unit{\mu s}$), the Medium Energy X-ray telescope
(ME, $5-30\unit{keV}$, $952\unit{cm}^{2}$, $1^{\degree}\times 4^{\degree}$, $%
\sim 276\unit{\mu s}$), and the Low Energy X-ray telescope (LE, $1-15\unit{%
keV}$, $384\unit{cm}^{2}$, $1.6^{\degree}\times 6^{\degree}$, $\sim 1\unit{ms%
}$) \citep{Zhang2014}. The exposure is about 520\thinspace ks for
observation P0101297 and P0101322. The data reduction for PSR B0540-69
observations is done by HXMTDAS software v2.0 and the data processing is
described in \cite{Chen2018}, \cite{Huang2018} and \cite{Tuo19}.

NICER is an X-ray instrument mounted on a movable arm on the outside of the
International Space Station (ISS) and has been in operation since June 2017 %
\citep{Gendreau2017}, which was specifically designed to study the X-ray
emissions of neutron stars. PSR B0540-69 is also the target of NICER. The
observation P102001 is used to analyze the timing properties. We only select
the photons within Good Time Intervals, which is generated by the following
four criteria: the ISS is not within the South Atlantic Anomaly ; NICER is
in tracking mode; NICER is pointing within 0.015\thinspace degrees of the
source; and the source is at least 30\thinspace degrees above the Earth's
limb.

\begin{table}
\caption{The calculated braking index of B0540--69 by using the data from
HXMT and NICER as well as the data from \textsl{NuSTAR}, \textsl{Swift/XRT}
and \textsl{XMM-Newton}.}
\label{tbl:n}
\begin{center}
\begin{tabular}{ccc}
\hline\hline
timing start (MJD) & timing end (MJD) & $n$ \\ \hline
57070 & 57303 & $0.11\pm 0.09$ \\ 
57331 & 57546 & $0.09\pm 0.2$ \\ 
57546 & 57743 & $0.2\pm 0.1$ \\ 
57755 & 57945 & $0.1\pm 0.2$ \\ 
57950 & 58129 & $0.5\pm 0.3$ \\ 
58160 & 58419 & $0.8\pm 0.08$ \\ 
58562 & 58748 & $1.2\pm 0.2$ \\ \hline
\end{tabular}%
\end{center}
\end{table}

The calculation for time of arrival (ToA) and timing process could be found
in \cite{Ge2019}. In order to show the evolution of braking index $n$, we
divide the observations into seven parts (see Table \ref{tbl:n}). The timing
solution in every part is obtained from the full coherent timing analysis by
TEMPO2 \citep{Hobbs06}.

\section{The evolution of the braking index}

\label{sec:n}

For the idealized magnetic dipole model in a vacuum, the braking index is $%
n=3$ \citep{Ostriker69, Shapiro83}. However, the measured braking indices of
young pulsars usually differ from 3 
\citep[see the catalogue
in][]{Manchester2005}. To account for this, many models were proposed 
\citep{Gunn70, Macy74, Blandford83, Beskin84, Candy86, Melatos97, Menou01,
Bucciantini06, Contopoulos06, Li12, Lyne13, Liu14, Philippov14, Kou15,
Hamil2015, Eksi16, Tong2017, Gao17, Beskin2018,Petri2019}. More
specifically, the braking index can be smaller than 3 due to the changing
moment of inertia (MoI) of the neutron star \citep[e.g.,][]{Hamil2015}, the
evolution of inclination angle between the rotational axis and the magnetic
dipole axis \citep[e.g.,][]{Tong2017}, a monopolar component in the
relativistic magnetised wind \citep[e.g.,][]{Petri2019}, and/or the
evolution of surface magnetic field \citep[e.g.,][]{Eksi17}.

Usually the braking index of a pulsar keeps constant for a very long time.
This remains true for PSR B0540-69, which keeps $n=2.129\pm 0.012$ for a
long time. What makes B0540-69 unusual is that its braking index changed to $%
n=0.031\pm 0.013$ \citep{Marshall2016} within two weeks in Dec. 2011. This
cannot be caused by a sudden change of the angle between the rotational axis
and the magnetic dipole axis because such a change takes place only
gradually \citep{Michel70, Philippov14}. Furthermore, we notice that the
predicted braking index is $n\gtrsim 1$ for models such as the change of MoI 
\citep[Figs
3,4,9,10 in][]{Hamil2015}, the evolution of inclination angle 
\citep[Fig. 4
in][]{Tong2017}, and the effect of monopolar spindown component %
\citep[Equations 66 and 72 in][]{Petri2019}. The braking index of $%
n=0.031\pm 0.013$\ cannot be explained by these models. \cite{Eksi17}
interpreted the rapid change of the braking index of B0540-69 as the growth
of dipole fields submerged by initial accretion of fallback matter soon
after the supernova explosion that formed B0540-69 
\citep{Young95,Geppert99,
Espinoza11, Ho11, Vigano12, Bernal13, Guneydas13, Torres-Forne16}.

Here we suggest that the growth of dipole magnetic field is a plausible
interpretation of the braking index evolution and leave more discussion in
Section \ref{sec:dis}. The sudden change of the spin-down rate $\dot{\nu}$\
in Dec. 2011 indicates an increase of the magnetic field by $\Delta
B_{p}/B_{p}\approx 16.7\%$ (from $\sim 6\times 10^{12}\unit{G}$ to $\sim
7\times 10^{12}\unit{G}$)\ within two weeks. Thereafter the magnetic field
increases smoothly. In this scenario, the braking index is calculated as %
\citep[e.g.,][]{Gao17}%
\begin{equation}
n=3+2\frac{\dot{B}_{p}}{B_{p}}\frac{\Omega }{\dot{\Omega}}=3-\frac{\dot{B}%
_{p}}{B_{p}}\frac{12Ic^{3}}{B_{p}^{2}R_{\ast }^{6}\Omega ^{2}\sin ^{2}\alpha 
},
\end{equation}%
where $B_{p}$, $R_{\ast }$, $\alpha $ and $I$ are the dipole magnetic field,
radius, inclination angle and moment of inertia of the pulsar, respectively.
In the following calculation we set\ $\alpha =90^{\circ }$. Here, it
indicates that $n<3$ for $\dot{B}_{p}>0$, otherwise, $n\geq 3$. We adopt the
dipole field growth history by the following equation%
\begin{equation}
B_{p}=B_{p0}\left\{ 1+R\left[ 1-e^{-\left( t-t_{1}\right) /\tau _{B}}\right]
\right\} ,  \label{eq:B-evol}
\end{equation}%
where $\tau _{B}$ is the field growth timescale, $t_{1}$ the time when the
field begin to increase, $R$ the field growth ratio, and $B_{p0}$ is the
field strength before growth. As discussed in Section \ref{sec:dis}, the
magnetic field growth may be approximated by a power-law or exponential
function of time. To make the field growth rate smooth enough, here we
decide to use Equation $\left( \ref{eq:B-evol}\right) $\ as an
approximation. The evolution of the braking index from Dec. 2011 to Dec.
2016 can be approximately described by the above equation. From Dec. 2016
on, the braking index begins to increase rapidly. To follow this rapid
increase, we decide to use the same equation $\left( \ref{eq:B-evol}\right) $
but choose another group of fitting parameters.

\begin{figure}
\centering\includegraphics[width=0.48\textwidth]{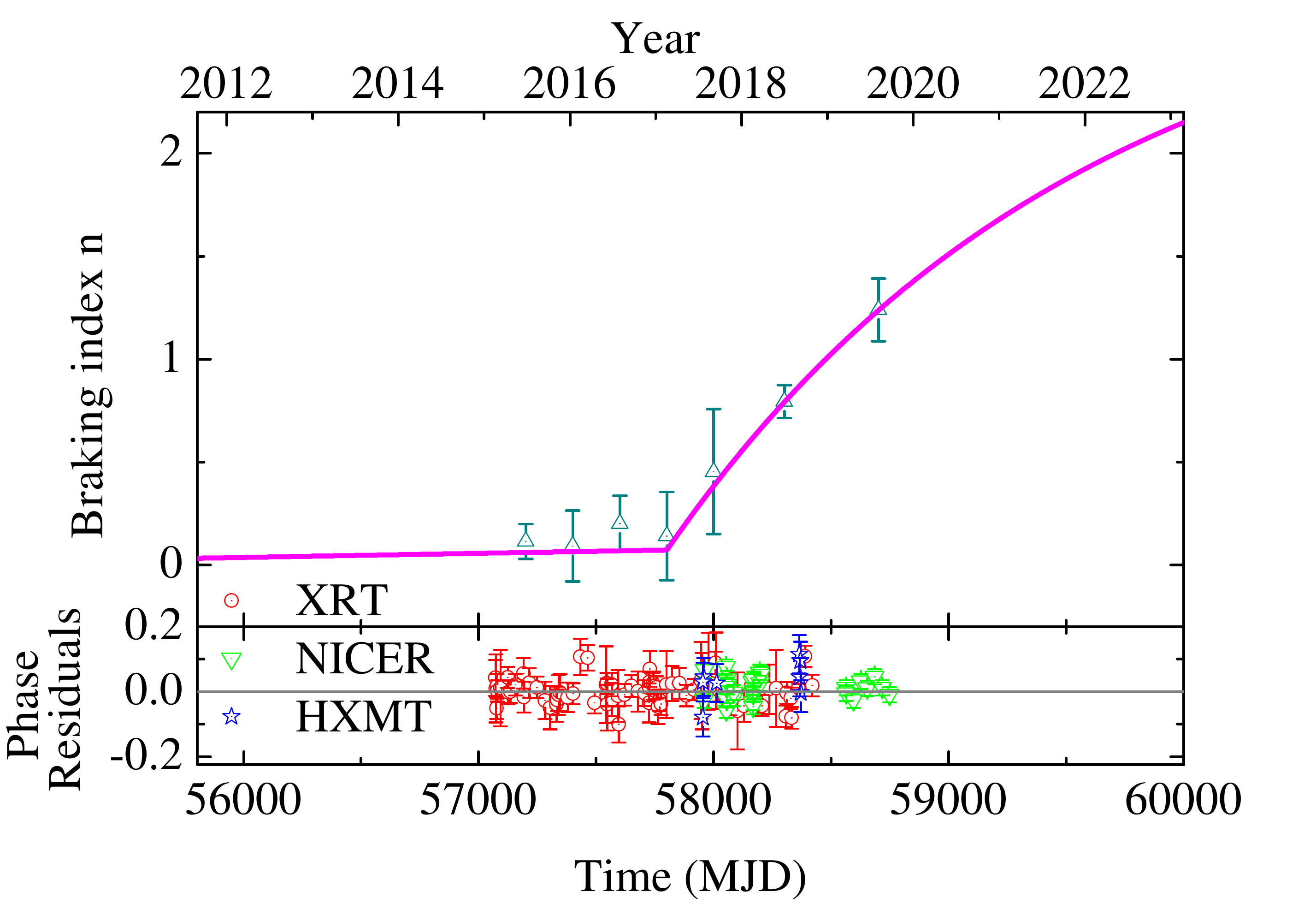}
\caption{\textsl{Top}: Braking index evolution since the SRT. The braking
index will restore to the pre-SRT value ($n=2.129$) in 2023$_{-2}^{+7}$ if
it continues to increase. The triangle data are newly calculated by
shortening the data process time interval. The new data allow us to follow
its evolution more closely. \textsl{Bottom}: timing residuals of XRT, NICER,
and HXMT.}
\label{fig:n}
\end{figure}

The braking index evolution is shown in Fig. \ref{fig:n}, and the fitting
parameters are presented in Table \ref{tbl:n-para}. The inclusion of the
data from Insight-HXMT and NICER confirms the evolution of $n$ and makes the
pulsar timing more accurate (see bottom panel of Fig. \ref{fig:n}). We can
see that this model fits the data very well. It also indicates that during
episode 1 (MJD 55800-57800) $\tau _{B}$ is very long, while this time scale
is very short in episode 2 (MJD 57800-).

\begin{table}
\caption{Best-fitting parameters for the evolution of braking index of
B0540--69.}
\label{tbl:n-para}
\begin{center}
\begin{tabular}{cccc}
\hline\hline
$R$ & $t_{1}$ (MJD) & $\tau _{B}$ (days) & $\frac{\Delta B_{p}}{B_{p}}$ \\ 
\hline
\multicolumn{4}{c}{episode 1} \\ 
$0.4_{-0.36}^{+0.8}$ & $55800\pm 7$ & $\left( 2.4_{-2.2}^{+5}\right) \times
10^{5}$ & $3.3\times 10^{-3}$ \\ \hline
\multicolumn{4}{c}{episode 2} \\ 
$\left( 2.9_{-0.9}^{+2.8}\right) \times 10^{-3}$ & $57800_{-260}^{+160}$ & $%
1800_{-600}^{+2600}$ & $1.4\times 10^{-3}$ \\ \hline
\end{tabular}%
\end{center}
\par
\textbf{Notes. }$t_{1}$ is the beginning time of the episode. $\tau _{B}$ is
the timescale of the magnetic increase. $\Delta B_{p}/B_{p}$ is the increase
ratio of the dipole magnetic field during the episode. $\Delta B_{p}/B_{p}$
is calculated according to the best-fit values and no fitting uncertainty is
provided here. For \textquotedblleft episode 2" $\Delta B_{p}/B_{p}$ is the
increase ratio from MJD 57800 to MJD 59000.
\end{table}

\section{The PWN brightening}

\label{sec:PWN}

The spin-down rate transition enhanced the energy injection rate to the PWN
and resulted in PWN X-ray brightening. The pulsar wind is highly magnetised
close to the magnetosphere and then quickly becomes lepton-dominated %
\citep{Kennel84a, Kennel84b, Aharonian12}, leading to a synchrotron PWN.

We first derive the equation that governs the PWN luminosity evolution.
Energy conservation yields%
\begin{equation}
L_{\mathrm{sd}}-N_{e}P_{\mathrm{syn}}-P\frac{dV}{dt}=\frac{dE}{dt},
\label{eq:energy-cons1}
\end{equation}%
where $L_{\mathrm{sd}}$ is the pulsar spin-down luminosity, $N_{e}$ is the
total electron (including positron) number capable of emitting synchrotron
photons, $P_{\mathrm{syn}}$ is the synchrotron power of one electron. $%
E=N_{e}E_{e}+E_{B}$ is the total energy contained in the PWN, including the
total electron energy $N_{e}E_{e}$ and magnetic energy in the PWN. $E_{e}$
is the average electron energy, which is calculated as%
\begin{equation}
E_{e}=\left( \frac{4\upi m_{e}c\nu }{3q_{e}B_{\mathrm{PWN}}}\right)
^{1/2}m_{e}c^{2}.  \label{eq:E_e}
\end{equation}%
Here $B_{\mathrm{PWN}}$ is the magnetic field in PWN, $\nu $ is the
frequency of the measured X-ray emission from the PWN, $m_{e}$ and $q_{e}$
are the mass and charge of an electron, respectively. The third term in
Equation $\left( \ref{eq:energy-cons1}\right) $ is the energy loss due to
volume expansion. Here we ignore the leakage of electrons and positrons from
the PWN. This is valid because PSR B0540-69 is surrounded by its nebula,
which confines electrons and positrons within the PWN %
\citep[e.g.,][]{Hooper09}. Equation $\left( \ref{eq:energy-cons1}\right) $
assumes that all the spin-down energy goes into the PWN, but some of that
energy powers the electromagnetic radiation seen from the pulsar. We ignore
this pulsed emission since the pulsar's X-ray and $\gamma $-ray luminosity $%
L_{X+\gamma }\sim 9.7\times 10^{36}\unit{erg}\unit{s}^{-1}$\ \citep{Fermi15}
is $\sim 0.06$\ of the pulsar's spin-down luminosity. Because the volume
expansion timescale is much longer than the synchrotron emission timescale%
\footnote{%
The stored energy in the PWN surrounding PSR B0540-69 is lost mainly through
X-ray emission. Therefore the synchrotron emission timescale of the
electrons is the lifetime of the electrons emitting X-ray.}, the third term
in the above equation can be neglected. Therefore the energy conservation
equation can be written as%
\begin{equation}
L_{\mathrm{sd}}-N_{e}P_{\mathrm{syn}}=\frac{dE}{dt}.  \label{eq:energy-cons}
\end{equation}

An electron with Lorentz factor $\gamma $ will produce synchrotron photons
with characteristic frequency \citep{Rybicki79}%
\begin{equation}
\nu =\frac{3\gamma ^{2}q_{e}B_{\mathrm{PWN}}}{4\upi m_{e}c}.
\end{equation}%
The synchrotron power of an electron with Lorentz factor $\gamma $ is %
\citep{Rybicki79}%
\begin{equation}
P_{\mathrm{syn}}=\frac{4}{3}\sigma _{T}c\gamma ^{2}\frac{B_{\mathrm{PWN}}^{2}%
}{8\upi },
\end{equation}%
where $\sigma _{T}$ is the Thomson cross section. The PWN X-ray luminosity is%
\begin{equation}
L_{X}=fN_{e}P_{\mathrm{syn}},  \label{eq:LX-P-syn}
\end{equation}%
where $f<1$ denotes the fraction of X-ray luminosity over the total
synchrotron power. The synchrotron emission timescale is%
\begin{equation}
\tau _{\mathrm{syn}}=\frac{\gamma m_{e}c^{2}}{P_{\mathrm{syn}}},
\label{eq:tau_syn}
\end{equation}%
from which the PWN magnetic field can be solved%
\begin{equation}
B_{\mathrm{PWN}}=\left( \frac{27\upi m_{e}cq_{e}}{\sigma _{T}^{2}\nu \tau _{%
\mathrm{syn}}^{2}}\right) ^{1/3}.
\end{equation}

\begin{figure}
\centering\includegraphics[width=0.48\textwidth]{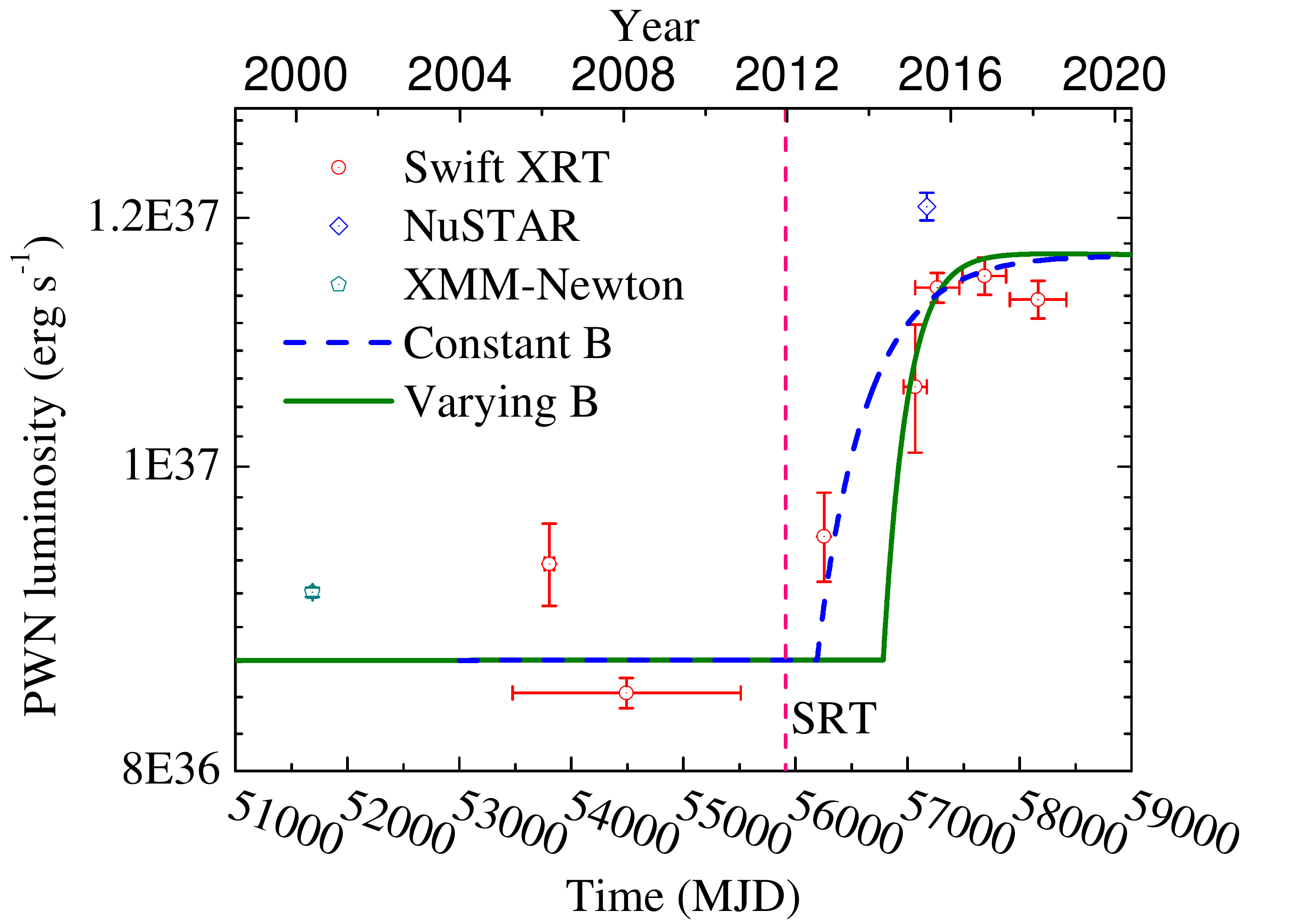}
\caption{X-ray luminosity of the PWN. The vertical dashed line marks the
time of SRT. The blue dashed line is the theoretical PWN X-ray luminosity
calculated assuming that the PWN magnetic field keeps constant even as the
energy injection rate from the pulsar increased. The solid green line is the
theoretical PWN X-ray luminosity calculated according to energy
equipartition.}
\label{fig:L-PWN}
\end{figure}

First we assume that the PWN magnetic field is constant before and after
SRT, then according to Equations $\left( \ref{eq:E_e}\right) $\ and $\left( %
\ref{eq:tau_syn}\right) $, $E_{e}$\ and $\tau _{\mathrm{syn}}$\ are
constants. In this case Equation $\left( \ref{eq:energy-cons}\right) $ is
reduced to%
\begin{equation}
\frac{L_{\mathrm{sd}}}{E_{e}}-\frac{N_{e}}{\tau _{\mathrm{syn}}}=\frac{dN_{e}%
}{dt}.  \label{eq:Ne-evolution}
\end{equation}%
Because the spin-down power of B0540-69 after SRT is approximately constant,
the above equation can be integrated to yield \citep{Ge2019}%
\begin{equation}
L_{X}=L_{X0}\left[ 1+\epsilon \left( 1-e^{-\left( t-t_{1}\right) /\tau _{%
\mathrm{syn}}}\right) \right] ,
\end{equation}%
where $\epsilon =\left( L_{X,\mathrm{new}}-L_{X0}\right) /L_{X0}$, $L_{X0}$
is the PWN X-ray luminosity before SRT, $L_{X,\mathrm{new}}$ is the new
steady X-ray luminosity after SRT, and $t_{1}$ is the time when the SRT
occurred.

In Fig. \ref{fig:L-PWN} we show the fitting result assuming a constant PWN
magnetic field as dashed curve (hereafter constant-$B$\ model). We find a
time delay between the SRT and PWN X-ray brightening $\sim 390$\ days,
indicating a termination shock at a distance of $R_{\mathrm{TS}}=c\Delta
t=0.3$\ parsec from the central pulsar, where we take into account the fact
that the relativistic particles in the pulsar wind move at a speed close to
the speed of light.

Theoretical studies 
\citep{Fried59, Weibel59, Davidson72, Wallace87, Wallace91, Yang94,
Califano98, Kazimura98, Medvedev99} imply that the magnetic energy density
in the PWN is proportional to the kinetic energy density of the electrons,
that is, $B_{\mathrm{PWN}}^{2}/8\upi =\eta _{B}N_{e}E_{e}/V$, $V$ is volume
of the PWN. In this case Equation $\left( \ref{eq:energy-cons}\right) $ can
be written as%
\begin{equation}
\frac{dN_{e}}{dt}=\left( 1+\eta _{B}\right) ^{-1}\left( \frac{L_{\mathrm{sd}}%
}{E_{e}}-\frac{N_{e}}{\tau _{\mathrm{syn}}}\right) -\frac{N_{e}}{E_{e}}\frac{%
dE_{e}}{dt}.
\end{equation}%
The last term takes into account the fact that the electrons emitting the
same X-ray photons in a stronger magnetic field need not to be as energetic
as the old ones in a weaker magnetic field. Usually $\eta _{B}\ll 1$, the
above equation can be approximated as%
\begin{equation}
\frac{dN_{e}}{dt}=\frac{L_{\mathrm{sd}}}{E_{e}}-\frac{N_{e}}{\tau _{\mathrm{%
syn}}}-\frac{N_{e}}{E_{e}}\frac{dE_{e}}{dt}.  \label{eq:Ne-varying-B}
\end{equation}

The PWN X-ray luminosity calculated according to $\left( \ref%
{eq:Ne-varying-B}\right) $ is presented in Fig. \ref{fig:L-PWN} as solid
line (hereafter varying-$B$\ model). In this model the time delay between
the SRT and PWN X-ray brightening is $\sim 980$ days, indicating a
termination shock at a distance of $R_{\mathrm{TS}}=c\Delta t=0.8$ parsec
from the central pulsar.

From Fig. \ref{fig:L-PWN} we see that both models fit the data well. The
difference between these two models is the rising time of the brightening.
Inspection of the observational data indicates that the PWN X-ray emission
at $\sim \mathrm{MJD}$ 56200\ was similar to that at $\sim \mathrm{MJD}$
53800\ and the brightening occurred at $\sim \mathrm{MJD}$ 57000. Based on
this fact, we see that the theoretical rising time of the constant-$B$ model
(dashed line) is longer than the observational data. This indicates that the
PWN magnetic field increased during the PWN X-ray brightening, resulting in
a shorter synchrotron timescale.\ The solid line can closely fit the rapid
rising (the data point at $\sim \mathrm{MJD}$ 57000) of the PWN X-ray
emission. To our knowledge, this is the first time that the time-resolved
link between the SRT and the enhancement of the PWN emission provides
observational evidence for magnetic field generation by a relativistic
shock. However, because of the sparse coverage of the observational data
during the rapid rise phase, it is premature to draw a solid conclusion on
the evidence of the generation of magnetic field by a relativistic shock.

From Equations $\left( \ref{eq:LX-P-syn}\right) $\ and $\left( \ref%
{eq:Ne-evolution}\right) $\ we can see that for a steady state (i.e., $%
dN_{e}/dt=0$) an approximate relation $L_{X}\propto N_{e}\propto L_{\mathrm{%
sd}}\propto B_{p}^{2}$\ is held, or equivalently 
\begin{equation}
\frac{\Delta L_{X}}{L_{X}}=2\frac{\Delta B_{p}}{B_{p}}.  \label{eq:delta_L_X}
\end{equation}%
Based on the increasing of dipole magnetic field model, the total magnetic
field growth from the SRT to the PWN brightening is $\Delta
B_{p}/B_{p}\approx 16.7\%$, while the observed X-ray luminosity change is ${%
\Delta L_{X}}/{L_{X}}\approx 32\%$. Therefore, the PWN X-ray flux can be
explained without other assumptions. This provide further evidence that the
PWN X-ray brightening was caused by the SRT.

Here we use a simple model with uniform properties throughout the PWN. Real
PWNe are complicated with, for example, large variations in the density and
average energies of energetic electrons 
\citep{Gaensler06,
Hester08, Kargaltsev15}. Large-scale variations are expected if the size of
the PWN is large compared with the termination shock radius %
\citep{Kennel84a, Kennel84b}. According to the X-ray observation, the PWN
associated with PSR B0540-69 has a spacial radius of $\sim 2.5^{\prime
\prime }$ (0.6\ parsec) in the sky in 2000 \citep{Gotthelf2000}. This radius
is similar to the termination shock radius. Therefore, the large-scale
variations are small and do not affect our conclusion. There are also
small-scale variations, which are caused by some instabilities, for
instance, the Rayleigh-Taylor, Kelvin-Helmholtz, or kink instabilities 
\citep{Chandrasekhar61, Hester96, Begelman98, Bucciantini04,
Bucciantini06a}. Such small-scale variations usually do not affect the
overall emission too much\footnote{%
Sometimes gamma-ray flares \citep{Abdo11, Tavani11, Buehler12} may occur in
PWNe that exceed the quiet emission in the same band, but such flares have
not been observed in X-ray bands. The peak luminosity of the gamma-ray
flares was $2\times 10^{36}\unit{erg}\unit{s}^{-1}$, about 1\% of the
spin-down power of the pulsar \citep{Abdo11, Tavani11, Porth17}. Therefore
the gamma-ray flares do not influence the total emission too much.}.

In this calculation we assume that the PWN is roughly spherical and the
electrons are relativistic. In this case the delays in the light travel
times from different parts of the PWN will modify the light curve (Fig. \ref%
{fig:L-PWN}). During the rising phase of the PWN light curve, we will first
receive emission from the nearest part of the PWN. Only after time $\Delta
t_{\mathrm{cross}}=2R_{\mathrm{TS}}/c$\ can we receive emission from the
farthest part of the PWN. This analysis indicates that the rising of the PWN
brightening should be slower than the light curves presented in Fig. \ref%
{fig:L-PWN}. After $\Delta t_{\mathrm{cross}}$, the emission should be close
to the theoretical prediction. This indicates that the results reproduced by
this simple model are reliable as a first approximation.

\section{Discussion}

\label{sec:dis}

In this paper, by including the data from HXMT and NICER to reduce the
timing noise, we calculate the braking index of PSR B0540-69 with shorter
timescale so that we can follow its evolution more closely. The braking
index of PSR B0540-69 evolved from $n\simeq 0.07$\ to\ $n\approx 1$\ in
about 2 years. We find that most of the current models can only explain the
braking index of $n\geq 1$. Therefore, these models cannot give an
acceptable explanation to the observed braking index evolution of PSR
B0540-69. One exceptional model is that the increase of the dipole magnetic
field of the pulsar. In this scenario the braking index evolution can be
well explained. In addition, the PWN X-ray luminosity is expected to
increase according to Equation $\left( \ref{eq:delta_L_X}\right) $, and this
is consistent with follow-up observations.

\cite{Eksi17} suggests that the small braking index $n=0.03$\ of PSR
B0540-69 is caused by the growth of dipole fields submerged by initial
fallback accretion. This is supported by the observed relation between the
measured velocities and the characteristic magnetic field growth timescale
of the pulsars. A nascent pulsar with a smaller kick velocity would accrete
more amount of matter, resulting in a deeper burial of its magnetic field
and a less quick field growth. PSR B0540-69 is moving fast and therefore the
burial of its magnetic field is shallow and the magnetic field grows quickly.

The short timescale (no longer than two weeks) of the SRT seems to suggest
an abrupt change of PSR B0540-69 in Dec. 2011. This may be caused by local
fractures in the crust of the pulsar. As the magnetic field diffuses out of
the neutron star core, small-scale fractures may be generated by the motion
of the magnetic field lines around the neutron star surface %
\citep{Thompson96, Mereghetti08}. This may suggest that the pre-SRT braking
index $n=2.12$\ is also caused by the growth of the dipole magnetic field.

Besides the field growth after the submergence by initial fallback accretion %
\citep{Young95}, we point out that the surface dipole magnetic field of a
neutron star may grow even without the initial submergence. Large-scale
magnetic field rearrangement is not rare in newborn neutron stars %
\citep{Bonanno03} and during the giant flares of magnetars %
\citep{Thompson96, Geppert06}. Because of some instabilities within the
neutron star, such as Tayler instability, slow and local magnetic field
rearrangements occur more frequently 
\citep{Chandrasekhar53, Ferraro54, Roxburgh63, Monaghan65, Tomimura05, Haskell08, 
Kiuchi08, Ciolfi09, Ciolfi11}. For a magnetic field configuration in a
neutron star in which the toroidal component is much stronger than the
poloidal one, field rearrangements may result in a configuration with
similar toroidal and poloidal components. In this case the surface dipole
field of a neutron star may grow as a result of such field rearrangements.
The fact that most of the young pulsars have braking indices $n<3$\ seems to
suggest that such field rearrangements may also occur within these pulsars.

The aforementioned surface field growth may occur even when the interior
magnetic field decays in a neutron star \citep{Goldreich92}. Alternatively,
the magnetic field in a neutron star may be enhanced by magnetorotational
instability \citep{Balbus91,Balbus98} and R-mode 
\citep{Andersson98, Rezzolla00,
Arras03, Wang17}, although these effects may be very small for a neutron
star as old as PSR B0540-69.

One puzzling fact of the observations of PSR B0540-69 and its PWN is that
despite the increase of the spin-down rate of the pulsar, the pulsed X-ray
emission did not change significantly \citep[$<10$\%;][]{Ge2019}. It is
therefore speculated that the change in the pulsar is mainly in the magnetic
polar region. If this is true, it implies that the magnetic field after the
SRT is not strictly dipole. In this case the surface field around the
fractures may be many times larger than the field without the fractures.
Assuming that all of the above speculation is correct, the fractures restore
to the pre-SRT state in about five years.

Theoretical studies \citep{Fried59, Weibel59, Medvedev99} and numerical
simulations \citep{Frederiksen04, Jaroschek04, Jaroschek05} of the
generation of magnetic field by relativistic shock has been carried out for
a long time. The SRT of PSR B0540-69 and the associated PWN brightening
provide us some evidence that the magnetic field in the PWN associated with
PSR B0540-69 was generated by the termination shock. However, a solid
conclusion can only be drawn if future high cadence observations of PWN
brightenings associated with SRTs are available.

\section*{Acknowledgements}

We thank Fang-Jun Lu for constructive comments on the manuscript. This work
is supported by the National Key Research and Development Program of China
(2016YFA0400800) and the National Natural Science Foundation of China under
grants 11673006, 11673013, 11653004, U1838201, U1838202, U1838104, and
U1938109. JSW is partially supported by China Postdoctoral Science
Foundation.
\section*{Note added in proof}
After the acceptance of the manuscript, the papers about ambipolar diffusion
in neutron stars \citep{Goldreich92, Thompson96, Beloborodov16, Wang20} come
into our attention. For the internal magnetic field of PSR B0540-69 to
diffuse out to the neutron star surface within 1100 year, an internal
magnetic field of $B_{\ast }\approx 4\times 10^{15}(t/1100\unit{yr})^{-0.83}%
\unit{G}$ is required.

\bsp
\label{lastpage}


\begin{thebibliography}{G\"{u}neyda\c{s} \& Ek\c{s}i(2013)}
\bibitem[Abdo et al.(2011)]{Abdo11} Abdo A. A. et al., 2011, Science, 331,
739

\bibitem[Aharonian et al.(2012)]{Aharonian12} Aharonian F. A., Bogovalov S.
V., Khangulyan D., 2012, Nature, \bibinfo{volume}{482}, \bibinfo{pages}{507}

\bibitem[Andersson(1998)]{Andersson98} Andersson N., 1998, ApJ, 502, 708

\bibitem[Archibald et al.(2016)]{Archibald16} Archibald R. F., Gotthelf E.
V., Ferdman R. D. et al., 2016, ApJ, 819, L16

\bibitem[Arons \& Scharlemann(1979)]{Arons79} Arons J., Scharlemann E. T.,
1979, ApJ, 231, 854

\bibitem[Arras et al.(2003)]{Arras03} Arras P. et al., 2003, ApJ, 591, 1129

\bibitem[Balbus \& Hawley(1991)]{Balbus91} Balbus S. A., Hawley J. F., 1991,
ApJ, 376, 214

\bibitem[Balbus \& Hawley(1998)]{Balbus98} Balbus S. A., Hawley J. F., 1998,
Rev. Mod. Phys., 70, 1

\bibitem[Begelman(1998)]{Begelman98} Begelman M. C., 1998, ApJ, 493, 291
\bibitem[Beloborodov \& Li(2016)]{Beloborodov16} Beloborodov A. M., Li X.,
2016, ApJ, 833, 261

\bibitem[Bernal et al.(2013)]{Bernal13} Bernal C. G., Page D., Lee W. H.,
2013, ApJ, 770, 106

\bibitem[Beskin et al.(1984)]{Beskin84} Beskin V. S., Gurevich A. V.,
Istomin Ia. N., 1984, Ap\&SS, 102, 301

\bibitem[Beskin(2018)]{Beskin2018} Beskin V.~S.,\ 2018, Physics Uspekhi, 61,
353

\bibitem[Blandford et al.(1983)]{Blandford83} Blandford R. D., Applegate J.
H., Hernquist L., 1983, MNRAS, 204, 1025

\bibitem[Bonanno et al.(2003)]{Bonanno03} Bonanno A., Rezzolla L., Urpin V.,
2003, A\&A, 410, L33

\bibitem[Bucciantini et al.(2004)]{Bucciantini04} Bucciantini N., Amato E.,
Bandiera R., Blondin J. M., Del Zanna L., 2004, A\&A, 423, 253

\bibitem[Buehler et al.(2012)]{Buehler12} Buehler R. et al., 2012, ApJ, 749,
26

\bibitem[Bucciantini \& Del Zanna(2006)]{Bucciantini06a} Bucciantini N., Del
Zanna L., 2006, A\&A, 454, 393

\bibitem[Bucciantini et al.(2006)]{Bucciantini06} 
\bibinfo{author}{Bucciantini N., Thompson T. A., Arons J., Quataert E.,
Del Zanna L.,} 2006, \mnras, \bibinfo{volume}{368}, \bibinfo{pages}{1717}

\bibitem[Bucciantini et al.(2011)]{Bucciantini11} %
\bibinfo{author}{Bucciantini N., Arons J., Amato E.,} 2011, \mnras, %
\bibinfo{volume}{410}, \bibinfo{pages}{381}

\bibitem[Califano et al.(1998)]{Califano98} Califano F., Pegorano F.,
Bupanov S. V., Mangeney A., 1998, Phys. Rev. E, 57, 7048

\bibitem[Candy \& Blair(1986)]{Candy86} Candy B. N., Blair D. G., 1986, ApJ,
307, 535

\bibitem[Chandrasekhar \& Fermi(1953)]{Chandrasekhar53} Chandrasekhar S.,
Fermi E., 1953, ApJ, 118, 116

\bibitem[Chandrasekhar(1961)]{Chandrasekhar61} Chandrasekhar S., 1961,
Hydrodynamic and Hydromagnetic Stability. Oxford: Oxford Univ. Press

\bibitem[Chen et al.(2018)]{Chen2018} {Chen} Y.~P., {Zhang} S., {Qu} J.~L.
et al., 2018, ApJ, 864, L30

\bibitem[Ciolfi et al.(2009)]{Ciolfi09} Ciolfi R., Ferrari V., Gualtieri L.,
Pons J. A., 2009, MNRAS, 397, 913

\bibitem[Ciolfi et al.(2011)]{Ciolfi11} Ciolfi R., Lander S. K., Manca G.
M., Rezzolla L., 2011, ApJ, 736, L6

\bibitem[Contopoulos \& Spitkovsky(2006)]{Contopoulos06} Contopoulos I.,
Spitkovsky A., 2006, ApJ, 643, 1139

\bibitem[Coroniti(1990)]{Coroniti90} Coroniti F. V., 1990, ApJ, 349, 538

\bibitem[Dai(2004)]{Dai04} Dai Z. G., 2004, ApJ, 606, 1000

\bibitem[Davidson et al.(1972)]{Davidson72} Davidson R. C., Hammer D. A.,
Haber I., Wagner C. E., 1972, Phys. Fluids, 15, 317

\bibitem[Ek\c{s}i (2017)]{Eksi17} \bibinfo{author}{Ek\c{s}i K. Y.,} 2017, %
\mnras, \bibinfo{volume}{469}, \bibinfo{pages}{1974}

\bibitem[Ek\c{s}i et al.(2016)]{Eksi16} 
\bibinfo{author}{Ek\c{s}i K. Y., et
al.,} 2016, \apj, \bibinfo{volume}{823}, \bibinfo{pages}{34}

\bibitem[Espinoza et al.(2011a)]{Espinoza11} Espinoza C. M., Lyne A. G.,
Kramer M. et al., 2011a, ApJ, 741, L13

\bibitem[Espinoza et al.(2011b)]{Espinoza11b} Espinoza C. M., Lyne A. G.,
Stappers B. W., Kramer M., 2011b, MNRAS, 414, 1679

\bibitem[Ferdman et al.(2015)]{Ferdman15} Ferdman R. D., Archibald R. F.,
Kaspi V. M., 2015, ApJ, 812, 95

\bibitem[Fermi LAT Collaboration(2015)]{Fermi15} The Fermi LAT
Collaboration, 2015, Science, 350, 801

\bibitem[Ferraro(1954)]{Ferraro54} Ferraro V. C. A., 1954, ApJ, 119, 407

\bibitem[Frederiksen et al.(2004)]{Frederiksen04} Frederiksen J. T., Hededal
C. B., Haugb\o lle T., Nordlund \AA ., 2004, ApJ, 608, L13

\bibitem[Fried(1959)]{Fried59} Fried B. D., 1959, Phys. Fluids, 2, 337

\bibitem[Gaensler \& Slane(2006)]{Gaensler06} Gaensler B. M., Slane P. O.,
2006, ARA\&A, 44, 17

\bibitem[Gao et al.(2017)]{Gao17} 
\bibinfo{author}{Gao Z. F., Wang N.,
Shan H., Li X. D., Wang W.,} 2017, \apj, \bibinfo{volume}{849}, %
\bibinfo{pages}{19}

\bibitem[Ge et al.(2019)]{Ge2019} 
\bibinfo{author}{{Ge} M. Y., {Lu} F.
J., {Yan} L.~L., et al.,} 2019, Nature Astronomy, 3, 1122

\bibitem[Gendreau \& Arzoumanian(2017)]{Gendreau2017} Gendreau K.,
Arzoumanian Z., 2017, Nature Astronomy, 1, 895

\bibitem[Geppert et al.(1999)]{Geppert99} 
\bibinfo{author}{Geppert U.,
Page D., Zannias T.,} 1999, A\&A, \bibinfo{volume}{345}, \bibinfo{pages}{847}

\bibitem[Geppert \& Rheinhardt(2006)]{Geppert06} Geppert U., Rheinhardt M.,
2006, A\&A, 456, 639

\bibitem[Goldreich \& Reisenegger(1992)]{Goldreich92} Goldreich P.,
Reisenegger A., 1992, ApJ, 395, 250

\bibitem[Gotthelf \& Wang(2000)]{Gotthelf2000} {{Gotthelf} E.~V., {Wang}
Q.~D.,} 2000, ApJ, \bibinfo{volume}{532}, \bibinfo{pages}{L117}

\bibitem[G\"{u}neyda\c{s} \& Ek\c{s}i(2013)]{Guneydas13} G\"{u}neyda\c{s}
A., Ek\c{s}i K. Y., 2013, MNRAS, 430, L59

\bibitem[Gunn \& Ostriker(1970)]{Gunn70} Gunn J. E., Ostriker J. P., 1970,
ApJ, 160, 979

\bibitem[Hamil et al.(2015)]{Hamil2015} Hamil O., Stone J.~R., Urbanec M. et
al.,\ 2015, \prd, 91, 063007

\bibitem[Haskell et al.(2008)]{Haskell08} Haskell B., Samuelsson L.,
Glampedakis K., Andersson N., 2008, MNRAS, 385, 531

\bibitem[Hester(2008)]{Hester08} Hester J. J., 2008, ARA\&A, 46, 127

\bibitem[Hester et al.(1996)]{Hester96} Hester J. J. et al., 1996, ApJ, 456,
225

\bibitem[Ho(2011)]{Ho11} Ho W. C. G., 2011, MNRAS, 414, 2567

\bibitem[Hobbs et al.(2006)]{Hobbs06} Hobbs G. B., Edwards R. T., Manchester
R. N., 2006, MNRAS, 369, 655

\bibitem[Hooper et al.(2009)]{Hooper09} Hooper D., Blasi P., Serpico P. D.,
2009, JCAP, 01, 025

\bibitem[Huang et al.(2018)]{Huang2018} {Huang} Y., {Qu} J.~L., {Zhang}
S.~N. et al., 2018, \apj, 866, 122

\bibitem[Jaroschek et al.(2004)]{Jaroschek04} Jaroschek C. H., Lesch H.,
Treumann R. A., 2004, ApJ, 616, 1065

\bibitem[Jaroschek et al.(2005)]{Jaroschek05} Jaroschek C. H., Lesch H.,
Treumann R. A., 2005, ApJ, 618, 822

\bibitem[Kargaltsev et al.(2015)]{Kargaltsev15} Kargaltsev O., Cerutti B.,
Lyubarsky Y., Striani E., 2015, SSRv, 191, 391

\bibitem[Kazimura et al.(1998)]{Kazimura98} Kazimura Y., Sakai J. I.,
Neubert T., Bulanov S. V., 1998, ApJ, 498, L183

\bibitem[Kennel \& Coroniti(1984a)]{Kennel84a} 
\bibinfo{author}{Kennel C.
F., Coroniti F. V.,} 1984a, \apj, \bibinfo{volume}{283}, \bibinfo{pages}{694}

\bibitem[Kennel \& Coroniti(1984b)]{Kennel84b} 
\bibinfo{author}{Kennel C.
F., Coroniti F. V.,} 1984b, \apj, \bibinfo{volume}{283}, \bibinfo{pages}{710}

\bibitem[Kiuchi \& Yoshida(2008)]{Kiuchi08} Kiuchi K., Yoshida S., 2008,
Phys. Rev. D, 78, 044045

\bibitem[Kou \& Tong(2015)]{Kou15} Kou F. F., Tong H., 2015, MNRAS, 450, 1990

\bibitem[Li et al.(2012)]{Li12} Li J., Spitkovsky A., Tchekhovskoy A., 2012,
ApJ, 746, 60

\bibitem[Liu et al.(2014)]{Liu14} Liu X. W., Xu R. X., Qiao G. J. et al.,
2014, Res. Astron. Astrophys., 14, 85

\bibitem[Lyne et al.(2013)]{Lyne13} Lyne A., Graham-Smith F., Weltevrede P.
et al., 2013, Science, 342, 598

\bibitem[Lyne et al.(2015)]{Lyne15} Lyne A. G., Jordan C. A., Graham-Smith
F. et al., 2015, MNRAS, 446, 857

\bibitem[Macy(1974)]{Macy74} Macy W. W. Jr., 1974, ApJ, 190, 153

\bibitem[Manchester et al.(2005)]{Manchester2005} Manchester R.~N., Hobbs
G.~B., Teoh A. et al.,\ 2005, \aj, 129, 1993

\bibitem[Mathewson et al.(1980)]{Mathewson1980} Mathewson D.~S., Dopita
M.~A., Tuohy I.~R., Ford V.~L., 1980, ApJ, 242, L73

\bibitem[Marshall et al.(2015)]{Marshall2015} {{Marshall} F.~E., {Guillemot}
L., {Harding} A.~K., {Martin} P., {Smith} D.~A.,} 2015, ApJ, %
\bibinfo{volume}{807}, \bibinfo{pages}{L27}

\bibitem[Marshall et al.(2016)]{Marshall2016} 
\bibinfo{author}{{Marshall}
F.~E., {Guillemot} L., {Harding} A.~K., {Martin} P., {Smith}
D.~A.,} 2016, ApJ, \bibinfo{volume}{827}, \bibinfo{pages}{L39}

\bibitem[Medvedev \& Loeb(1999)]{Medvedev99} Medvedev M. V., Loeb A., 1999,
ApJ, 526, 697

\bibitem[Melatos(1997)]{Melatos97} Melatos A., 1997, MNRAS, 288, 1049

\bibitem[Menou et al.(2001)]{Menou01} Menou K., Perna R., Hernquist L.,
2001, ApJ, 554, L63

\bibitem[Mereghetti(2008)]{Mereghetti08} Mereghetti S., 2008, A\&ARv, 15, 225

\bibitem[Michel(1994)]{Michel94} Michel F. C., 1994, ApJ, 431, 397

\bibitem[Michel \& Goldwire(1970)]{Michel70} Michel F. C., Goldwire H. C.
Jr, 1970, Astrophys. Lett., 5, 21

\bibitem[Monaghan(1965)]{Monaghan65} Monaghan J. J., 1965, MNRAS, 131, 105

\bibitem[Ostriker \& Gunn(1969)]{Ostriker69} Ostriker J. P., Gunn J. E.,
1969, ApJ, 157, 1395

\bibitem[Ou et al.(2016)]{Ou16} Ou Z. W., Tong H., Kou F. F., Ding G. Q.,
2016, MNRAS, 457, 3922

\bibitem[Pacini \& Salvati(1973)]{Pacini1973} {Pacini} F., {Salvati} M.,
1973, ApJ, 186, 249

\bibitem[P{\'{e}}tri(2019)]{Petri2019} P{\'{e}}tri J.,\ 2019, \mnras, 485,
4573

\bibitem[Philippov et al.(2014)]{Philippov14} Philippov A., Tchekhovskoy A.,
Li J. G., 2014, MNRAS, 441, 1879

\bibitem[Porth et al.(2017)]{Porth17} Porth O. et al., 2017, SSRv, 207, 137

\bibitem[Rees \& Gunn(1974)]{Rees1974} {Rees} M.~J., {Gunn} J.~E., 1974,
MNRAS, 167, 1

\bibitem[Rezzolla et al.(2000)]{Rezzolla00} Rezzolla L., Lamb F. K., Shapiro
S. L., 2000, ApJ, 531, L139

\bibitem[Roxburgh(1963)]{Roxburgh63} Roxburgh I. W., 1963, MNRAS, 126, 67

\bibitem[Rybicki \& Lightman(1979)]{Rybicki79} Rybicki G. B., Lightman A.
P., 1979, Radiative Processes in Astrophysics (New York: Wiley-Interscience)

\bibitem[Shapiro \& Teukolsky(1983)]{Shapiro83} Shapiro S. L., Teukolsky S.
A., 1983, Black Holes, White Dwarfs, and Neutron Stars: The Physics of
Compact Objects (New York: Wiley)

\bibitem[Seward et al.(1984)]{Seward1984} {Seward} F.~D., {Harnden} Jr.
F.~R., {Helfand} D.~J., 1984, ApJ, 287, L19

\bibitem[Tavani et al.(2011)]{Tavani11} Tavani M. et al., 2011, Science,
331, 736

\bibitem[Thompson \& Duncan(1996)]{Thompson96} Thompson C., Duncan R. C.,
1996, ApJ, 473, 322

\bibitem[Tomimura \& Eriguchi(2005)]{Tomimura05} Tomimura Y., Eriguchi Y.,
2005, MNRAS, 359, 1117

\bibitem[Tong \& Kou(2017)]{Tong2017} Tong H., Kou F.~F.,\ 2017, \apj, 837,
117

\bibitem[Torres-Forn\'{e} et al.(2016)]{Torres-Forne16} Torres-Forn\'{e} A.,
Cerd\'{a}-Dur\'{a}n P., Pons J. A., Font J. A., 2016, MNRAS, 456, 3813

\bibitem[Tuo et al.(2019)]{Tuo19} Tuo Y. L., Ge M. Y., Song L. M., Yan L.
L., Bu Q. C., Qu J. L., 2019, Res. Astron. Astrophys., 19, 87

\bibitem[Vigan\`{o} \& Pons(2012)]{Vigano12} Vigan\`{o} D., Pons J. A.,
2012, MNRAS, 425, 2487

\bibitem[Wallace et al(1987)]{Wallace87} Wallace J. M. et al, 1987, Phys.
Fluids, 30, 1085

\bibitem[Wallace \& Epperlein(1991)]{Wallace91} Wallace J. M., Epperlein E.
M., 1991, Phys. Fluids B, 3, 1579

\bibitem[Wang \& Dai(2013)]{Wang13} Wang L. J., Dai Z. G., 2013, ApJ, 774,
L33
\bibitem[Wang \& Dai(2017)]{Wang17} Wang J. S., Dai Z. G., 2017, A\&A, 603,
A9

\bibitem[Wang \& Lai(2020)]{Wang20} Wang J. S., Lai D., 2020, ApJ, 892, 135

\bibitem[Wang et al.(2016)]{Wang16a} 
\bibinfo{author}{Wang L. J., Dai Z.
G., Liu L. D., Wu X. F.,} 2016, \apj, \bibinfo{volume}{823}, %
\bibinfo{pages}{15}

\bibitem[Weibel(1959)]{Weibel59} Weibel E. S., 1959, Phys. Rev. Lett., 2, 83

\bibitem[Yang et al.(1994)]{Yang94} Yang T.-Y., Arons J., Langdon A. B.,
1994, Phys. Plasmas, 1, 3059

\bibitem[Young \& Chanmugam(1995)]{Young95} 
\bibinfo{author}{Young E. J.,
Chanmugam G.,} 1995, ApJ, \bibinfo{volume}{442}, \bibinfo{pages}{L53}

\bibitem[Zhang et al.(2014)]{Zhang2014} Zhang S., Lu F. J., Zhang S. N., Li
T. P., 2014, in Takahashi T., den Herder J.-W. A., Bautz M., eds, Proc SPIE
Conf. Ser. Vol. 9144, Space Telescopes and Instrumentation 2014: Ultraviolet
to Gamma Ray. SPIE, Bellingham, p. 914421
\end{thebibliography}
\end{document}